\documentclass[10pt,a4paper,twocolumn, twoside]{article}
\usepackage{graphicx}
\usepackage[T1]{fontenc}
\usepackage[latin1]{inputenc}
\usepackage{amsmath}
\usepackage{caption}
\usepackage{booktabs}
\usepackage{setspace}
\usepackage{etoolbox}
\usepackage{amssymb}

\usepackage[dvipsnames]{xcolor}

\AtBeginEnvironment{tabular}{\onehalfspacing}

\makeatletter
\renewcommand{\section}{\@startsection{section}{2}{0cm}{-\baselineskip}
{0,5\baselineskip}{\normalsize\bfseries}}
\renewcommand{\subsection}{\@startsection{subsection}{3}{0cm}{-\baselineskip}
{0,5\baselineskip}{\normalsize\slshape}}
\makeatother

\RequirePackage{mathptmx}      
\RequirePackage[numbers,sort&compress]{natbib}
\RequirePackage[colorlinks,citecolor=blue,urlcolor=blue,linkcolor=blue]{hyperref}
\RequirePackage{siunitx}
\DeclareSIUnit[]\ppb{ppb}
\DeclareSIUnit[]\ppt{ppt}
\DeclareSIUnit[]\ppq{ppq}
\DeclareSIUnit[]\ccm{\cubic\cm}
\DeclareSIUnit[]\mbar{\milli\bar}

\begin{document}

\title{Radon depletion in xenon boil-off gas}

\author{S. Bruenner, D. Cichon, S. Lindemann, T. Marrod\'an Undagoitia, H. Simgen}

\date{\small \it
Max-Planck-Institut f\"ur Kernphysik, Saupfercheckweg 1, D-69117 Heidelberg, Germany \\
\vspace{0.3cm}
Email-address: \\ {\tt stefan.bruenner@mpi-hd.mpg.de} 
}

\twocolumn[
\begin{@twocolumnfalse}
\maketitle
\begin{abstract}
\noindent	An important background in detectors using liquid xenon for rare event searches arises from the decays of radon and its daughters. We report for the first time a reduction of $^{222}$Rn in the gas phase above a liquid xenon reservoir. 
We show a reduction factor of $\gtrsim 4$ for the $^{222}$Rn concentration in boil-off xenon gas compared to the radon enriched liquid phase. A semiconductor-based $\alpha$-detector and miniaturized proportional counters are used to detect the radon. 
As the radon depletion in the boil-off gas is understood as a single-stage distillation process, this result establishes the suitability of cryogenic distillation to separate radon from xenon down to the $10^{-15}$\,mol/mol level.
\vspace{0.5cm}
\end{abstract}
\end{@twocolumnfalse}
]

\section{Introduction}
The radioactive noble gas $^{222}$Rn is known to be a crucial background source in rare event experiments using a liquid xenon target \cite{r1,r2,r3,r15,r24}. This radon isotope is continuously produced in all detector materials containing traces of $^{238}$U. Released by recoil or diffusion, $^{222}$Rn distributes inside the liquid xenon where it disintegrates.
The beta decay of $^{214}$Pb, which is part of the $^{222}$Rn decay chain, is indeed one of the most important backgrounds in dark matter experiments \cite{r15,r16,r17} and will become even more important for current and future ton-scale detectors \cite{r1, r18, r19}.
In order to reduce this internal source of background, large efforts are carried out to carefully preselect materials by demanding stringent limits on their radon emanation rates \cite{r20}.\\
However, once the detectors are constructed, the emanation rate and thus the $^{222}$Rn induced background is set at a certain level. As the isotope is dispersed throughout the liquid xenon target, its contribution to the background cannot be reduced by making use of self-shielding capabilities as it is the case for surface backgrounds.
A further mitigation of this background can be achieved by continuously removing radon from the xenon target, e.g. in a closed purification loop.
First tests using an adsorption based radon removal have been reported in \cite{r5,r20}.
Cryogenic distillation is an alternative technique commonly used to purify xenon from traces of krypton \cite{r6,r22,r23}.\\
In this paper, we demonstrate for the first time the capability to remove $^{222}$Rn from xenon below the $10^{-15}\,$mol/mol level by means of cryogenic distillation. In order to realize a single distillation stage, we liquefy radon enriched xenon in a dedicated system. The reduction in the $^{222}$Rn activity concentration in the gas phase is then quantified using both, a semiconductor-based $\alpha$-detector (radon monitor) and miniaturized proportional counters. In this work we focus only on $^{222}$Rn and refer to this isotope as radon.

\section{Experimental setup and measuring process}
\label{sec:setup}
\begin{figure*}[phtb]
	\centering
	\includegraphics[width=14cm]{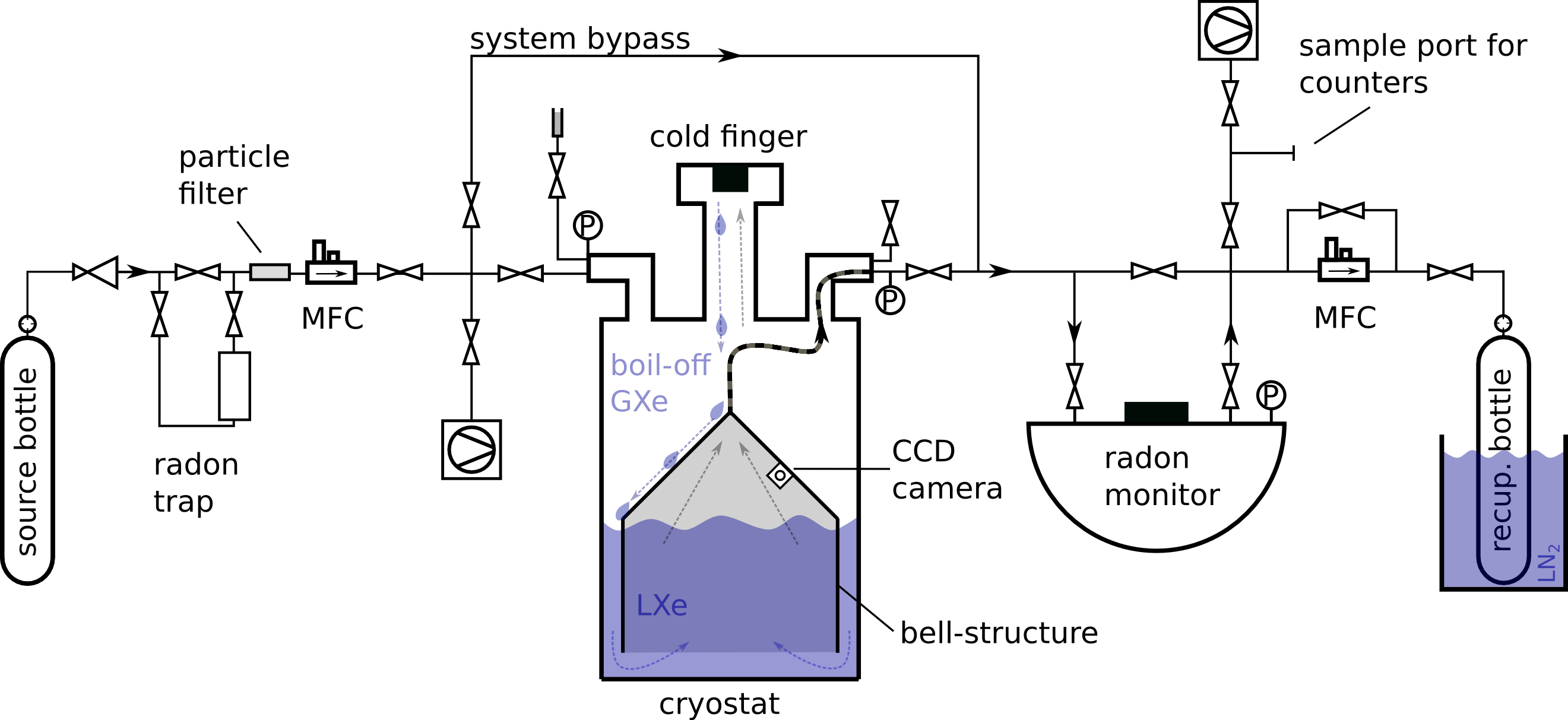}	
	\caption{\textsl{Experimental setup for measuring radon reduction in boil-off xenon. Central element is the cryostat for liquefaction and storage of xenon. After filling, 
	the boil-off gas above the liquid surface is recuperated via a radon monitor to measure its activity concentration. A bell structure inside the cryostat separates the  gas inlet from the outlet. 
	Sample ports allow to draw gas samples for independent measurements with proportional counters.}}
	\label{fig:setup}
\end{figure*}

A schematic of the experimental setup is shown in Fig.~\ref{fig:setup}. The central element is a vacuum insulated stainless steel cryostat which houses the radon enriched liquid xenon (LXe) reservoir. 
During the measurement, xenon from the gas phase above the liquid is cryogenically pumped from the cryostat through a radon monitor into a bottle immersed in liquid nitrogen (recup. bottle). In the following we refer to this operation as recuperation. The radon monitor facilitates continuous monitoring of the radon concentration in the gas phase and allows us to determine the reduction factor with respect to the liquid. Complementary measurements are done using miniaturized proportional counters. In this section, we describe the devices employed for radon detection and the procedure used in order to initially enrich the xenon reservoir with radon. The experimental setup and the process of measuring the boil-off reduction are discussed afterwards.

\subsection{Radon detection}
\label{subsect:det}
In order to measure the radon activity in the xenon gas, we use two different types of detectors. A radon monitor allows for a continuous on-line measurement of radon concentrations in the boil-off gas. These measurements are complemented using well characterized proportional counters in order to analyze gas samples drawn during the measuring process.\\ The radon monitor used in this work is an electrostatic ion-drift chamber with a Si-PIN photo diode as sensitive detector for $\alpha$-particles. Similar detectors are widely used in many experiments \cite{r7,r8,r9,r10}.
\begin{figure}[phtb]
	\centering
	\includegraphics[width=7.5cm]{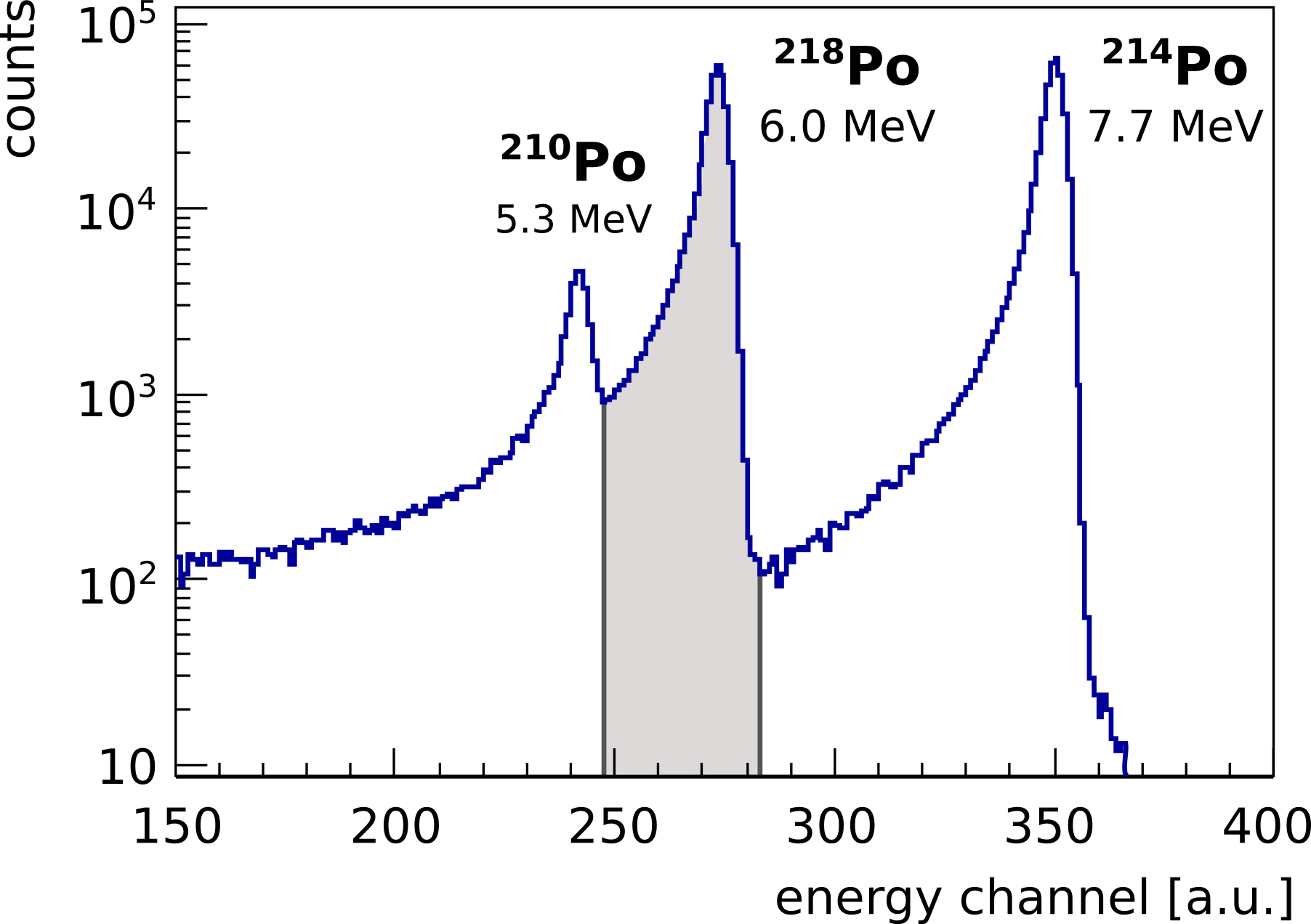}	
	\caption{\textsl{Detection of $\alpha$-decaying $^{222}$Rn progenies with the radon monitor. The gray area marks the selection window of $^{218}$Po events in our analysis.}}
	\label{fig:peaks}
\end{figure}
In our radon monitor, a Hamamatsu Si-PIN is placed inside a 1 liter stainless steel vessel. A high voltage of $-1.8$\,kV is applied to the diode with respect to the vessel's grounded walls. 
Thus, the positively charged progenies of $\alpha$-emitters such as $^{222}$Rn drift along the electric field lines onto the surface of the PIN diode. Subsequent $\alpha$-decays, e.g., $^{218}$Po and $^{216}$Po in the decay 
chain of $^{222}$Rn, are detected with an energy resolution of $<4$\% (Fig.~\ref{fig:peaks}). The tails of the peaks towards lower energies are due to events that do not deposit their full energy into the diode's active volume. Since $^{222}$Rn is not detected directly, the subsequent daughter isotope  $^{218}$Po, that has a rather short half-life of 3.1\,min, is used in the analyses to monitor the radon activity inside the radon monitor. 
The detection efficiency of the $^{218}$Po decay is $\epsilon_{218}=(0.222\pm0.005)$, determined using calibrated $^{222}$Rn standards. This yields an estimated sensitivity on activity concentrations of roughly $\sim$50\,Bq/kg at an operating pressure of 2.0\,bara xenon. Leakage from neighboring polonium peaks into the $^{218}$Po event selection window (gray area in Fig.~\ref{fig:peaks}) was found to be $<1\%$ in all measurements and is thus neglected.
The radon monitor showed a dependence on the operating pressure. Most likely, the positively charged ions are drifting less efficiently towards the diode as the pressure inside the chamber increases. 
This effect causes a correction of the data of up to 20\% when we compare activity concentrations measured at different pressure conditions.
The radon concentration of the boil-off gas is measured on-line while being flushed through the radon monitor with mass flows up to 8.0 standard liters per minute (slpm). At flow rates bigger than $6$\,slpm, we observed a decrease in efficiency of about 10\% which we also correct our data for.\\
The miniaturized proportional counters used in this work have been originally developed for the GALLEX solar neutrino experiment \cite{r11}. They were furthermore used 
to measure traces  of noble gases, e.g.,  $^{133}$Xe in air samples \cite{r12} and to study the $^{222}$Rn emanation rate of detector materials with a sensitivity of 40\,$\mu$Bq \cite{r13}.
In order to determine the radon reduction in boil-off xenon, gas samples are taken from a sample port installed after the radon monitor (see Fig.~\ref{fig:setup}) and stored in glass vessels. 
A dedicated filling facility is used to transfer a calibrated amount of the xenon sample into the proportional counters \cite{r14}.

\subsection{Radon enrichment of xenon}
In this work, we use an aqueous 20\,kBq $^{226}$Ra standard as a radon source. We apply a small helium purge flow to transfer the $^{222}$Rn from the aqueous source to a portable silica gel trap immersed in liquid nitrogen. An additional trap, cooled down to 
$-25^\circ$C, is placed between the source and the silica gel trap to remove humidity. The radon loaded silica gel trap is then integrated in the experimental setup (see Fig.~\ref{fig:setup}) and heated to $180^\circ$C. 
Xenon is flushed from the source bottle through the hot silica gel trap into a recuperation bottle immersed in liquid nitrogen. The cryostat and the radon monitor are bypassed during this operation. After the xenon has been transfered, the recuperation bottle is closed and  warmed up. It now contains xenon gas that is homogeneously enriched with radon. In the following measurements this bottle is used as source bottle (see Fig.~\ref{fig:setup}). Its radon concentration is determined by expanding gas into the radon monitor and analyzing the observed $^{218}$Po decays (see section~\ref{sec:analysis}). A complementary result is obtained from the measurement using a proportional counter.

\subsection{Boil-off reduction measurement}
At the beginning of each run, the cryostat is evacuated and the source bottle containing xenon with a known radon activity concentration is connected (see Fig.~\ref{fig:setup}). We fill xenon gas from the source bottle into the cryostat, bypassing the silica gel trap. Inside the setup, the xenon liquefies  at a cold finger connected to the cryostat via a pipe of 60\,mm diameter. Heating cartridges, which partly compensate the provided cooling power, are installed to establish an equilibrium 
between liquefaction at the cold finger and inflowing as well as evaporating gas. After filling a liquid xenon mass of 2 to 4\,kg (see Table~\ref{tab:1} for the precise values for the different runs), we close the gas inlet and let the system stabilize.
Eight PT100 temperature sensors, mounted at different heights close to the cryostat's wall, monitor the temperature and indicate the liquid xenon level. In addition, the hight of the liquid surface is observed by a CCD camera mounted inside the setup at the cryostat's top flange.
During run1 and run2 (see Table~\ref{tab:1}), the port to recuperate the boil-off gas was connected to a cylindrically-shaped bell structure shown in Fig.~\ref{fig:setup}. As long as the liquid level exceeds a height of 10\,mm (corresponding to about $0.9$\,kg of xenon), the bell structure separates the boil-off gas, which is recuperated, from the gas phase which is in contact with the gas inlet and the cold finger. 
From the cryostat, the radon reduced xenon is transfered through the radon monitor into the recuperation bottle. The 
radon monitor is continuously measuring the radon activity concentration of the recuperated gas. In each run, we fix the recuperation flow (see Table~\ref{tab:1}). The pressures inside the cryostat and the radon monitor are monitored by three pressure gauges.
As discussed in section~\ref{sec:analysis}, we use the monitored activity concentration of the xenon gas to determine the radon reduction factor with respect to the activity concentration in the liquid reservoir.\\
For complementary measurements with proportional counters, a port right after the radon monitor allows to take gas samples. For each run, we draw pipettes from the source bottle and from the boil-off gas.
\begin{table*}[phtb]
    \centering
    \begin{tabular}{c c c c c}
        \hline
                          &\textbf{filled} &  \textbf{act. conc. of}  & \textbf{recuperation-} &\textbf{bell structure} \\
                          &\textbf{xenon mass [kg]}&\textbf{filled gas [kBq/kg]}& \textbf{flow [slpm]}&\textbf{present}\\ \hline
  \textbf{run1}       &$2.1\pm0.1$      &    $4.50 \pm0.09 $          &    0.5                          &yes\\
  \textbf{run2}       &$2.7\pm0.1$      &    $1.76\pm0.03$          &    0.55                         &yes\\
  \textbf{run3}       &$2.6\pm0.1$      &    $1.22\pm0.02$          &    0.5, 3.5                  &no\\
  \textbf{run4}       &$4.0\pm0.1$      &    $0.87\pm0.02$          &    0.5, 6.0                  &no\\ \hline
      \end{tabular}
  \caption{Overview of the four runs presented in this work. The runs differ not only by the xenon mass filled into the cryostat, but also in the presence resp. absence of the bell structure. In run1 and run2, the recuperation flow is kept constant 
  while in the other runs it was increased to a higher value after measurement at 0.5\,slpm.}
  \label{tab:1}
\end{table*}

\section{Data Analysis}
\label{sec:analysis}
\subsection{Radon monitor measurements}
\begin{figure}[phtb]
	\centering
	\includegraphics[width=8cm]{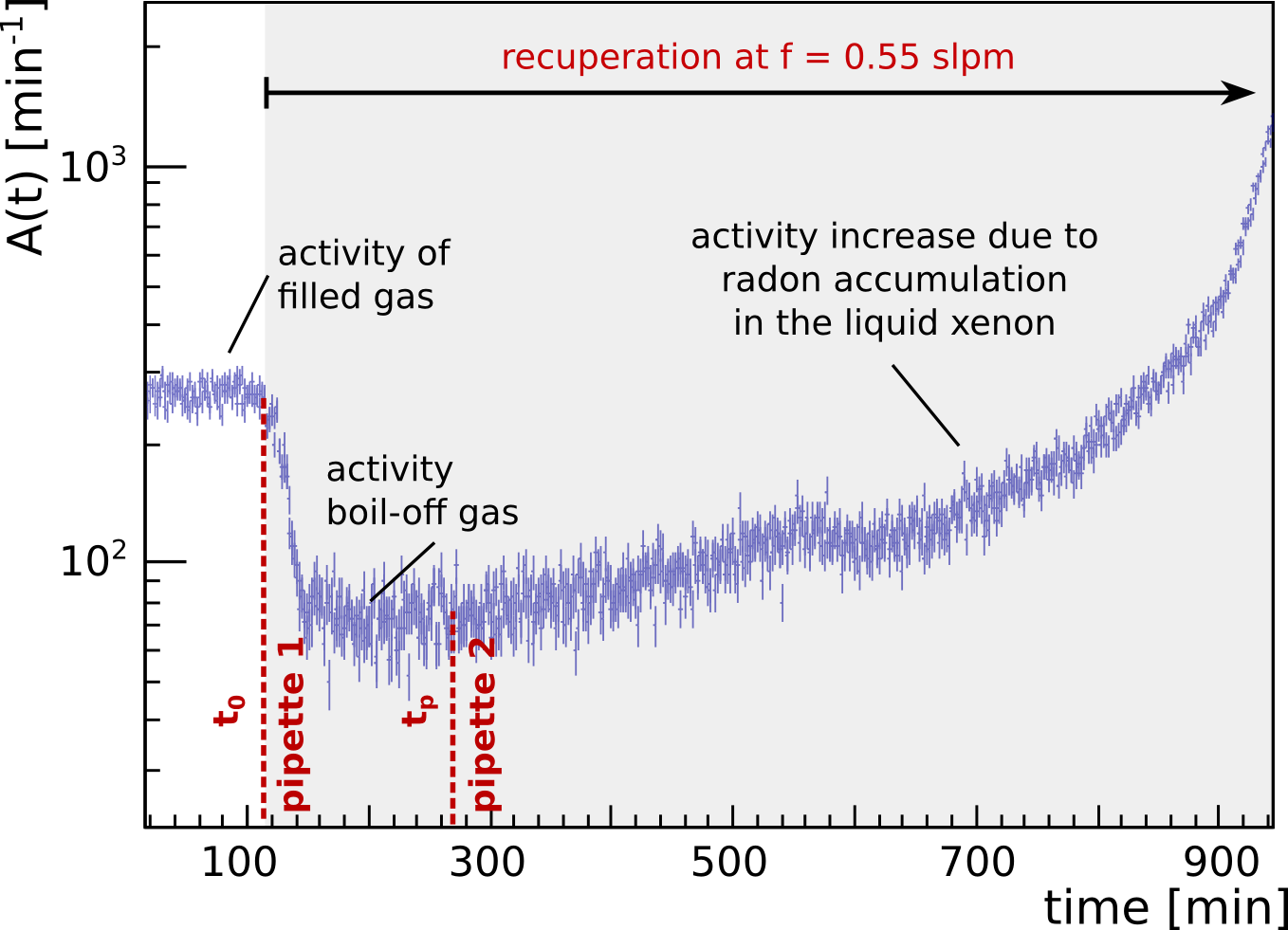}	
	\caption{\textsl{Radon monitor data of a typical run (run2). Two gas samples (pipette 1, pipette 2) are taken for complementary measurements with proportional counters. The observed increase of the monitored $^{222}$Rn activity in the course of the run is a clear signature for the accumulation of radon in the liquid phase.}}
	\label{fig:run4}
\end{figure}

Data from a typical run acquired with the radon monitor can be seen in Fig.~\ref{fig:run4} displaying the evolution of the activity in the boil-off gas with time. The high activity at the beginning of the measurement corresponds to the activity concentration of the radon enriched gas from the source bottle, $c_g(t_0)$. At about $t_0=110$\,min, we start the recuperation of the boil-off gas. The observed drop of the measured activity is a clear signature 
for radon reduction in the boil-off gas. As a consequence, radon accumulates in the liquid phase while we continuously empty the cryostat. This evolution also explains the increasing radon activity observed in the boil-off gas (see Fig.~\ref{fig:run4}).
In order to determine the boil-off reduction during each run, we define the reduction factor $R$ as
\begin{equation}
\centering
    R(t) \equiv \frac{c_l(t)}{c_g(t)} \; ,
    \label{eq:1}
\end{equation}
where $c_l(t)$ and $c_g(t)$ are the activity concentrations in units of Bq/kg of the liquid and boil-off xenon, respectively.
$c_g(t)$ is given by the ratio of the measured activity $A(t)$ and the amount of xenon inside the radon monitor at a certain time $t$. We can use the density dependent (see section~\ref{subsect:det}) detection efficiency $\epsilon_D(\rho_g^D(t))$, to write
\begin{equation}
 \centering
 c_g(t) = \frac{A(t)}{V_D\cdot \rho^D_{g}(t)\cdot \epsilon_D(\rho_g^D(t))} \cdot e^{\lambda_{Rn}\cdot(t-t_0)}\quad .
 \label{eq:2}
\end{equation}
Since $V_D$ denotes the volume of the radon monitor, the denominator in equation~\ref{eq:2} represents the xenon mass inside the detector. The exponential factor corrects for the radioactive decay of radon during the run with respect to the
recuperation start at $t=t_0$.
Analogously, we determine $c_l(t)$ by the ratio between the activity dissolved in the liquid phase $A_l(t)$ and the liquid xenon mass $M_l(t)$.
After introducing the total filled xenon mass $M_0$, and the monitored recuperation mass flow $f(t)$, we can write
\begin{equation}
 \centering
 M_l(t) = M_0-V_g^C(t)\cdot \rho^C_g(t) - \int_{t_0}^{t}f(t')\,dt' \; .
  \label{eq:3}
\end{equation}
The product of $V_g^C(t)$, i.e. the volume occupied by the boil-off gas inside the cryostat, and its density $\rho^C_g(t)$ gives the total mass of the gas phase inside the setup. In this analysis we neglect the time dependence of $V_g^C(t)$, caused by the 
decreasing liquid level inside the cryostat. Its impact on $M_l(t)$ was found to be $<1\%$ in all runs.\\
The total activity inside the liquid phase can be written as
\begin{equation}
 \centering
A_l(t) = A_0-c_g(t)\cdot V_g^C\cdot \rho^C_g(t) - \int_{t_0}^t c_g(t')\cdot f(t')\, dt' \; .
\label{eq:4}
\end{equation}
Here, $A_0$ is the total filled activity given by $A_0=c_g(t_0)\cdot M_0$, while $c_g(t_0)$ is the activity concentration of the initial radon enriched gas from the source bottle (equation~\ref{eq:2}).
The activity concentration in the liquid xenon reservoir is then calculated as the ratio between equations \ref{eq:4} and \ref{eq:3} as
\begin{equation}
 \centering
 c_l(t)=\frac{A_l(t)}{M_l(t)} \; .
 \label{eq:4a}
\end{equation}
Using the definition in equation~\ref{eq:1}, we compute
\begin{align}
\label{eq:ramonR}
& R(t)\equiv \frac{c_l(t)}{c_g(t)} =  \\ 
&= \frac{c_g(t_0)\cdot M_0 - c_g(t)\cdot V_g^C\cdot \rho^C_g(t)-\int_{t_0}^t c_g(t')\cdot f(t')\,dt'}{c_g(t)\cdot \left(M_0-V_g^C \cdot \rho^C_g(t) - \int_{t_0}^t f(t')\, dt'\right)} \; .\nonumber
\end{align}
In section~\ref{sec:results}, equation~\ref{eq:ramonR} is used to investigate the evolution of the boil-off reduction factor $R$ during the different runs.
\subsection{Proportional counter measurements}
Measurements with the proportional counters do not allow for a continuous monitoring of the boil-off reduction factor. Instead, $R(t_p)$ is determined only at a certain time $t_p$ during a run (see Fig.~\ref{fig:run4}). 
In this analysis, we consider $t_p$ to be close enough to the recuperation start $t_0$ such that we can assume $c_l(t_p)=c_l(t_0)$, i.e. the radon enrichment in the liquid is negligible at $t=t_p$. The liquid concentration is then determined by evaluating equations \ref{eq:3} and \ref{eq:4} at recuperation start
\begin{equation}
 \centering
c_l(t_p) =  \frac{c_g^p(t_0)\cdot M_0-c_g^p(t_p)\cdot V_g^C\cdot \rho_g^C(t_p)}{M_0-V_g^C\cdot \rho_g^C(t_p)} \; .
\label{eq:5}
\end{equation}
$c_g^p(t_0)$ and $c_g^p(t_p)$ are the activity concentrations obtained with the proportional counters for pipette 1 and pipette 2 respectively (see Fig.~\ref{fig:run4}).
The boil-off reduction factor is then calculated as
\begin{equation}
 \centering
R(t_p) =  \frac{c_g(t_0)\cdot M_0-c_g(t_p)\cdot V_g^C\cdot \rho_g^C(t_p)}{c_g(t_p)\cdot \left(M_0-V_g^C\cdot \rho_g^C(t_p)\right)} \; .
\label{eq:p5}
\end{equation}

\section{Results and Discussion}
\label{sec:results}
\begin{figure}[phtb]
	\centering
	\includegraphics[width=8cm]{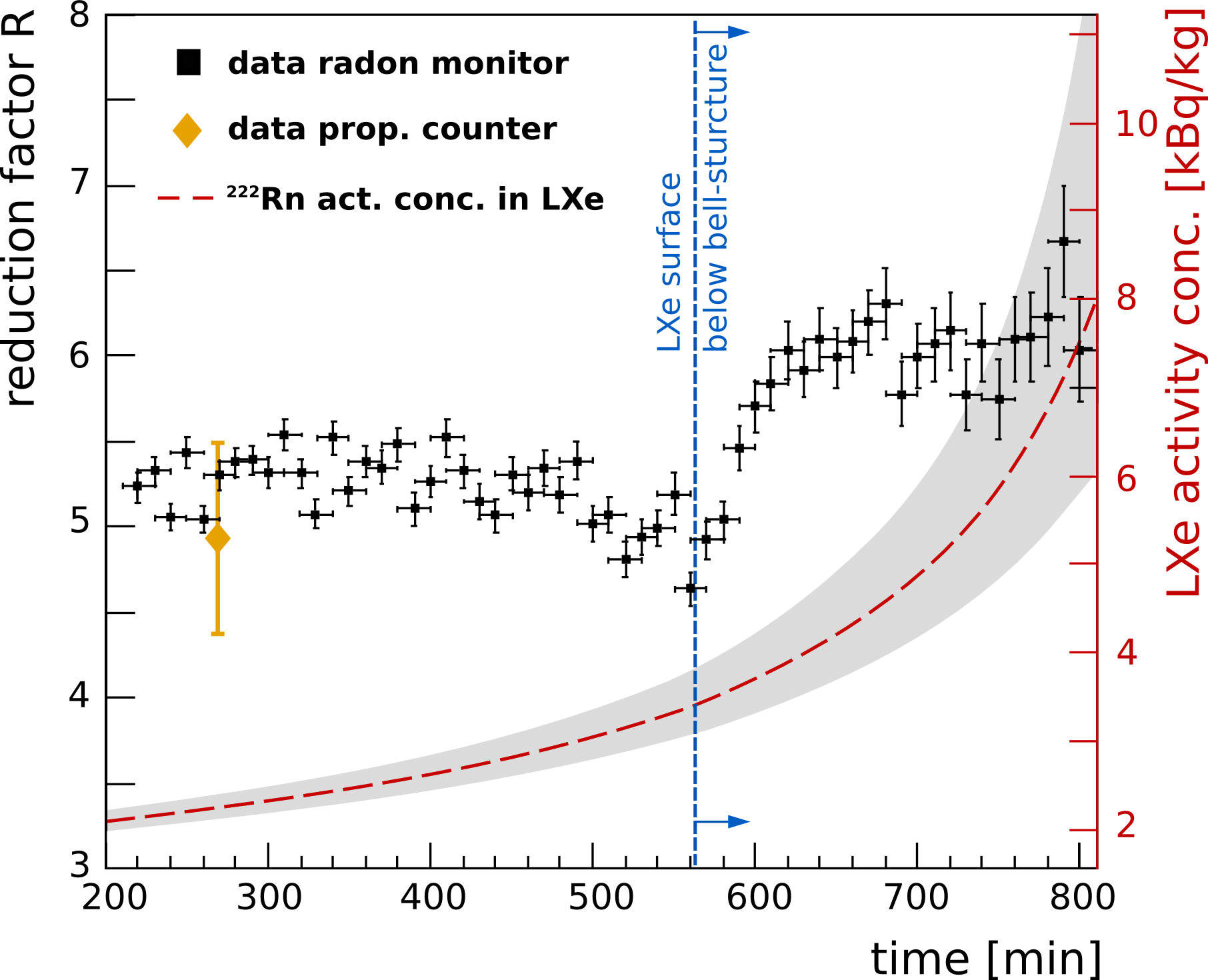}	
	\caption{\textsl{Measured radon reduction factor during run2. While $R$ is unaffected by the changing activity concentration $c_l(t)$, a clear increase is seen when the liquid xenon (LXe) surface falls underneath the bell structure.}}
	\label{fig:run4_result}
\end{figure}

During all four runs, we observe a similar evolution of the radon activity concentration. Therefore, we exemplary discuss run2. Table~\ref{tab:2} summarizes the results of all runs. The reduction factor $R$ as a function of time is shown in Fig.~\ref{fig:run4_result}. In run2, the liquid xenon reservoir has been 
recuperated with a constant mass flow of $0.55$\,slpm. Thus, the time axis correlates to the liquid xenon mass inside the cryostat. The increasing activity concentration in the liquid reservoir, $c_l(t)$, along with its uncertainty is shown in Fig.~\ref{fig:run4_result} 
with a dashed red line.\\ We clearly identify two plateau regions in the evolution of $R$ during this run. 
From $t = 200$\,min  until $t = 500$\,min, we find $R = (5.27 \pm0.05)$. This result is in good agreement with the 
complementary measurement using proportional counters (orange diamond at $t=250$\,min). The increase of the liquid xenon activity concentration $c_l(t)$ by a factor 1.5 within this time period shows no significant impact on the reduction factor. Instead, we find a clear signature in $R$ at $t=550$\,min, the time when the liquid xenon level drops below the bottom edge of the bell structure. After a small dip, the measured reduction factor increases to a constant value of $R=(6.0\pm0.1)$. This signature appears also in run1,
but is absent in run3 and run4 where no bell structure was installed.\\
In order to emphasize the impact of the bell structure, Fig.~\ref{fig:run4_result} only shows the statistical errors, which are dominated by the activity measurement with the radon monitor. The larger systematic errors, up to 20\%, are driven by the uncertainty on 
the mass flow and on the temperature of the boil-off gas inside the cryostat. The latter is used to determine the total mass of the gas phase inside the cryostat (see section~\ref{sec:analysis}). The errors are given explicitly in Table~\ref{tab:2}, together with the final results of all runs obtained for a recuperation flow in the range of $0.5$\,slpm to $0.6$\,slpm. Complementary measurements with the proportional counters are marked with an asterisk (*). In Table~\ref{tab:2}, we distinguish between the phases where the liquid level is above the bell structure and those where the liquid xenon surface drops below the bottom edge of the bell. The latter is treated equivalently to having no bell installed as it is the case in run3 and run4.
In all runs we find a clear radon reduction in the boil-off xenon with respect to the liquid reservoir. Variations in the activity concentration in the liquid xenon, $c_l(t)$, do not affect the obtained reduction factor. In all runs, the initial activity concentration is in the range of a few kBq/kg (see Table~\ref{tab:1}) and increases during the recuperation phase (see Fig.~\ref{fig:run4_result}). The effect of the bell structure on $R$ is clearly visible in the course of run1 and run2 and cannot be explained by systematic uncertainties.
The three complementary measurements with proportional counters confirm the results obtained with the radon monitor. Due to the counters small active volume of $\sim 1$\,cm$^3$, their activity measurements suffer from larger statistical errors. In run4, we measure the lowest reduction factor, being not in agreement with the other results obtained under similar conditions.\\
The observed impact of the bell structure is possibly caused by the smaller liquid surface confined by the bell structure where the recuperated xenon gas is evaporated from. Additional heat input by the bell structure might influence the measurement as well. Larger recuperation flows of up to 
6.0\,slpm, as in run3 and run4 (no bell structure), on the other hand do not impact the reduction factor which was found unchanged.
\\In all measurements summarized in Table~\ref{tab:2}, we find $R\gtrsim 4$. 
\begin{table}[phtb]
 \centering
 \begin{tabular}{c c c} \hline
                          &  \textbf{above} &  \textbf{below/no}\\
                          & \textbf{bell structure} & \textbf{bell structure}  \\\hline  
                          \noalign{\vskip 5pt} 
 \textbf{run1}        &$\;\;4.61 \pm 0.02\,\textrm{\tiny{stat}} \; ^{+0.29}_{-0.27}\,\textrm{\tiny{sys}}$         &   $\;\;5.58 \pm 0.02\,\textrm{\tiny{stat}} \; ^{+0.88}_{-0.68}\,\textrm{\tiny{sys}}$                        \\[5pt]
                          & ${\color[rgb]{0,0,0}^*3.75 \pm 0.50\,\textrm{\tiny{stat}} \; ^{+0.08}_{-0.06}\,\textrm{\tiny{sys}}}$  & - \\[10pt]
 \textbf{run2}        &$\;\;5.27 \pm 0.05\,\textrm{\tiny{stat}} \; ^{+0.22}_{-0.27}\,\textrm{\tiny{sys}}$                 &   $\;\;6.02 \pm 0.04\,\textrm{\tiny{stat}} \; ^{+1.12}_{-0.78}\,\textrm{\tiny{sys}}$                        \\[5pt]
                          & ${\color[rgb]{0,0,0}^*4.91 \pm 0.68\,\textrm{\tiny{stat}} \; ^{+0.07}_{-0.05}\,\textrm{\tiny{sys}}}$ & - \\[10pt]
 \textbf{run3}        &       -            &   $\;\;7.20 \pm 0.04\,\textrm{\tiny{stat}} \; ^{+0.50}_{-0.31}\,\textrm{\tiny{sys}}$                        \\[5pt]
                          & - & ${\color[rgb]{0,0,0}^*8.12 \pm 1.35\,\textrm{\tiny{stat}} \; ^{+0.13}_{-0.10}\,\textrm{\tiny{sys}}}$ \\[10pt]
 \textbf{run4}        &        -           &   $\;\;3.77 \pm 0.09\,\textrm{\tiny{stat}} \; ^{+0.12}_{-0.13}\,\textrm{\tiny{sys}}$                       \\\hline
 \end{tabular}
\caption{Results for the radon reduction factor $R$ obtained in four measurements at recuperation flows of $0.5$\,slpm to $0.6$\,slpm (see Table~\ref{tab:1}). Complementary measurements using proportional counters are marked with an asterisk (*).}
\label{tab:2}
\end{table}
The reduction factor in a single distillation stage can be theoretically derived from Raoult's law using vapor saturation pressure curves.
For the radon reduction measurements presented here, the comparison is challenging due to various reasons: The vapor saturation curve of radon is not precisely measured~\cite{r25}, the reduction factor might be different at the very low concentrations investigated in this work or impacted by the dynamic measurement procedure where we continuously recuperated the boil-off gas. 

\section{Conclusions}
Cryogenic distillation is a widely used separation technique to purify gases. In this paper, we prove its applicability to separate radon from xenon  at concentrations as low as $10^{-15}$\,mol/mol. From our results we conclude that the radon concentration in the boil-off gas above a liquid xenon reservoir is reduced by a factor $R\gtrsim4$ with respect to the liquid concentration. Our setup realizes a single distillation stage. A multi-stage distillation column, as it is used for krypton removal~\cite{r22}, is supposed to achieve a much larger radon reduction. For liquid xenon based experiments, dealing with radon as an internal background source, this technique can be used to construct a radon removal system to continuously purify the liquid xenon target.



\end{document}